Preprint typeset in JINST style - HYPER VERSION

# Description and commissioning of NEXT-MM prototype: first results from operation in a Xenon-Trimethylamine gas mixture

V. Álvarez,[a] F. Aznar,[b,c] F.I.G.M. Borges,[d] D. Calvet,[e] S. Cárcel,[a] J. Castel,[b,c] S. Cebrián,[b,c] A. Cervera,[a] C.A.N. Conde,[d] T. Dafni,[b,c] T.H.V.T. Dias,[d] J. Díaz,[a] F. Druillole,[e] M. Egorov,[f] R. Esteve,[g] P. Evtoukhovitch,[h] L.M.P. Fernandes,[d] P. Ferrario,[a] A.L. Ferreira,[i] E. Ferrer-Ribas,[e] E.D.C. Freitas,[d] V.M. Gehman,[f] A. Gil,[a] I. Giomataris,[e] A. Goldschmidt,[f] H. Gómez,[b,c*†] J.J. Gómez-Cadenas,[a‡] D. González-Díaz,[b,c] R.M. Gutiérrez,[j] J. Hauptman,[k] J.A. Hernando Morata,[l] D.C. Herrera,[b,c] F.J. Iguaz,[b,c] I.G. Irastorza,[b,c] M.A. Jinete,[j] L. Labarga,[m] A. Laing,[a] A. Le Coguie,[e] I. Liubarsky,[a] J.A.M. Lopes,[d] D. Lorca,[a] M. Losada,[j] G. Luzón,[b,c] A. Marí,[g] J. Martín-Albo,[a] A. Martínez,[a] G. Martínez-Lema,[l] T. Miller,[f] A. Moiseenko,[h] J.P. Mols,[e] F. Monrabal,[a] C.M.B. Monteiro,[d] F.J. Mora,[g] L.M Moutinho,[i] J. Muñoz Vidal,[a] H. Natal da Luz,[d] G. Navarro,[j] M. Nebot-Guinot,[a] D. Nygren,[f] C.A.B. Oliveira,[f] R. Palma,[n] J. Pérez,[o] J.L. Pérez Aparicio,[n] J. Renner,[f] L. Ripoll,[p] A. Rodríguez,[b,c] J. Rodríguez,[a] F.P. Santos,[d] J.M.F. dos Santos,[d] L. Segui,[b,c] L. Serra,[a] D. Shuman,[f] A. Simón,[a] C. Sofka,[q] M. Sorel,[a] J.F. Toledo,[g] A. Tomás,[b,c] J. Torrent,[p] Z. Tsamalaidze,[h] J.F.C.A. Veloso,[i] J.A. Villar,[b,c] R.C. Webb,[q] J.T. White[q§] and N. Yahlali[a]

[a]

*Instituto de Física Corpuscular (IFIC), CSIC & Universitat de València*
*Calle Catedrático José Beltrán, 2, 46980 Paterna, Valencia, Spain*
[b] *Laboratorio de Física Nuclear y Astropartículas, Universidad de Zaragoza*
*Calle Pedro Cerbuna 12, 50009 Zaragoza, Spain*
[c] *Laboratorio Subterráneo de Canfranc*
*Paseo de los Ayerbe s/n, 22880 Canfranc Estación, Huesca, Spain*
[d] *Departamento de Fisica, Universidade de Coimbra*
*Rua Larga, 3004-516 Coimbra, Portugal*
[e] *IRFU, Centre d'Études Nucléaires de Saclay (CEA-Saclay)*
*91191 Gif-sur-Yvette, France*
[f] *Lawrence Berkeley National Laboratory (LBNL)*
*1 Cyclotron Road, Berkeley, California 94720, USA*
[g] *Instituto de Instrumentación para Imagen Molecular (I3M), Universitat Politècnica de València*
*Camino de Vera, s/n, Edificio 8B, 46022 Valencia, Spain*
[h] *Joint Institute for Nuclear Research (JINR)*
*Joliot-Curie 6, 141980 Dubna, Russia*
[i] *Institute of Nanostructures, Nanomodelling and Nanofabrication (i3N), Universidade de Aveiro*
*Campus de Santiago, 3810-193 Aveiro, Portugal*
[j] *Centro de Investigaciones en Ciencias Básicas y Aplicadas, Universidad Antonio Nariño*
*Carretera 3 este No. 47A-15, Bogotá, Colombia*
[k] *Department of Physics and Astronomy, Iowa State University*
*12 Physics Hall, Ames, Iowa 50011-3160, USA*
[l] *Instituto Gallego de Física de Altas Energías (IGFAE), Univ. de Santiago de Compostela*
*Campus sur, Rúa Xosé María Suárez Núñez, s/n, 15782 Santiago de Compostela, Spain*
[m] *Departamento de Física Teórica, Universidad Autónoma de Madrid*
*Campus de Cantoblanco, 28049 Madrid, Spain*
[n] *Dpto. de Mecánica de Medios Continuos y Teoría de Estructuras, Univ. Politècnica de València*
*Camino de Vera, s/n, 46071 Valencia, Spain*
[o] *Instituto de Física Teórica (IFT), UAM/CSIC*
*Campus de Cantoblanco, 28049 Madrid, Spain*
[p] *Escola Politècnica Superior, Universitat de Girona*
*Av. Montilivi, s/n, 17071 Girona, Spain*
[q] *Department of Physics and Astronomy, Texas A&M University*
*College Station, Texas 77843-4242, USA*

*E-mail:* `gomez@lal.in2p3.fr`

ABSTRACT: A technical description of NEXT-MM and its commissioning and first performance is reported. Having an active volume of ∼35 cm drift × 28 cm diameter, it constitutes the largest Micromegas-read TPC operated in Xenon ever constructed, made by a sectorial arrangement of the 4 largest single wafers manufactured with the Microbulk technique to date. It is equipped with a suitably pixelized readout and with a sufficiently large sensitive volume (∼23 l) so as to contain long (∼20 cm) electron tracks. First results obtained at 1 bar for Xenon and trimethylamine (Xe-(2%)TMA) mixture are presented. The TPC can accurately reconstruct extended background tracks. An encouraging full-width half-maximum of 11.6 % was obtained for ∼ 29 keV gammas without resorting to any data post-processing.

*Corresponding author.

†Present address: Laboratoire de l'Accélérateur Linéaire (LAL). Centre Scientifique d'Orsay. Bâtiment 200 - BP 34. 91898 Orsay Cedex, France.

‡Spokesperson (gomez@mail.cern.ch).

§deceased

## Contents

## 1. Introduction

NEXT-100 will search for the neutrinoless double beta decay ($0\nu\beta\beta$) of $^{136}$Xe with a high pressure Time Projection Chamber (TPC) [1, 2]. Its technological baseline relies on the detection of secondary light multiplication, namely electroluminescence, allowing for near-intrinsic energy resolution. The concept has been proven in several technological demonstrators (NEXT-DEMO, NEXT-DBDM) [3, 4, 5]. During the R&D phase a third prototype, NEXT-MM, was built to test an alternative option based on charge readout with high-pixelized micro-pattern gas structures (Micromegas).

NEXT-MM is a TPC with $\sim$35 cm drift length and 28 cm diameter ($\sim$23 l), large enough to contain few cm electron tracks. The TPC readout uses high-end micro-pattern gaseous detectors (MPGDs). It is made by a sectorial arrangement of the 4 largest (to date) single wafers microprocessed via the novel Microbulk technique [6]. It is the largest Micromegas-read TPC operated with Xenon ever constructed.

NEXT-MM currently serves as a general-purpose platform to test novel gas mixtures related to possible NEXT-100 upgrades. The potential of Xenon-trimethylamine (TMA) mixtures has been presented in [7, 8, 9] and is actively pursued by the NEXT collaboration, [10, 11]. The anticipated Penning effect has been identified in [12], while the low-diffusion characteristics have been recently addressed in [11] for the first time.

This paper describes the commissioning of NEXT-MM and its first performance, obtained at 1bar for a Xe-(2%)TMA mixture. The structure of the paper is as follows: the main elements of the setup, like the gas system, field cage, feedthroughs, readout and data acquisition electronics, are described in Section 2. The operational tests developed and first data taken in Ar-(2%)i$C_4H_{10}$ to test the different types of Micromegas detectors used are presented in Section 3. First results with a Xe-(2%)TMA mixture, including examples of the registered tracks, are summarized in Section 4. Finally, prospects and conclusions are presented in Sections 5 and 6, respectively.

## 2. Description of the experimental setup

The elements of NEXT-MM have been designed to work at high pressure and having low outgassing rates to keep the gas as pure as possible. Materials have been selected taking into account radiopurity criteria [13, 14]. Note that Micromegas detectors are a radiopure solution compared with other sensors, as noted in [15].

### 2.1 Vessel

All the detector elements are held in an 73-liter vessel made of ASIS 304-L Stainless Steel welded using conventional TIG welding. It is composed by a cylindrical central body of 5 mm thickness with 396 mm diameter and 590 mm length as internal dimensions, and two flat caps. Both, the vessel body and the caps, have several feedthroughs of different sizes for the installation of the required equipment for the detector operation, like gas inlet and outlet, high voltage supply and detector readout (see Figures 1 and 2). All these feedthroughs, as well as the assembling of the caps to the vessel body, have been done by resorting to standard CF-F flanges and copper gaskets in order to assure the required vacuum and high pressure conditions.

### 2.2 Internal components

#### 2.2.1 Field cage

The field cage is composed of 34 copper rings with an internal diameter of 280 mm plus one copper disk of the same diameter as cathode that can hold a calibration source (Figure 3-left). All these elements are assembled together using 4 bars of PEEK with 10 mm of separation between them, having a cylindrical field cage of 280 mm diameter and 340 mm drift length. The readout plane is mechanically attached to the bottom cap through PEEK pillars while the field cage is hanging from PEEK bars screwed to the top cap. The voltage degradation along the copper rings is obtained using 10 M$\Omega$ resistors between each of them. The lowest ring is connected to an external variable resistor that allows voltage tuning in order to have a homogeneous electric field depending on the voltage applied to the cathode and to the Micromegas mesh. The upper two thirds of the field cage have been surrounded by a Cirlex screen as insulator, avoiding possible sparks between the rings and the vessel body (Figure 3-right).

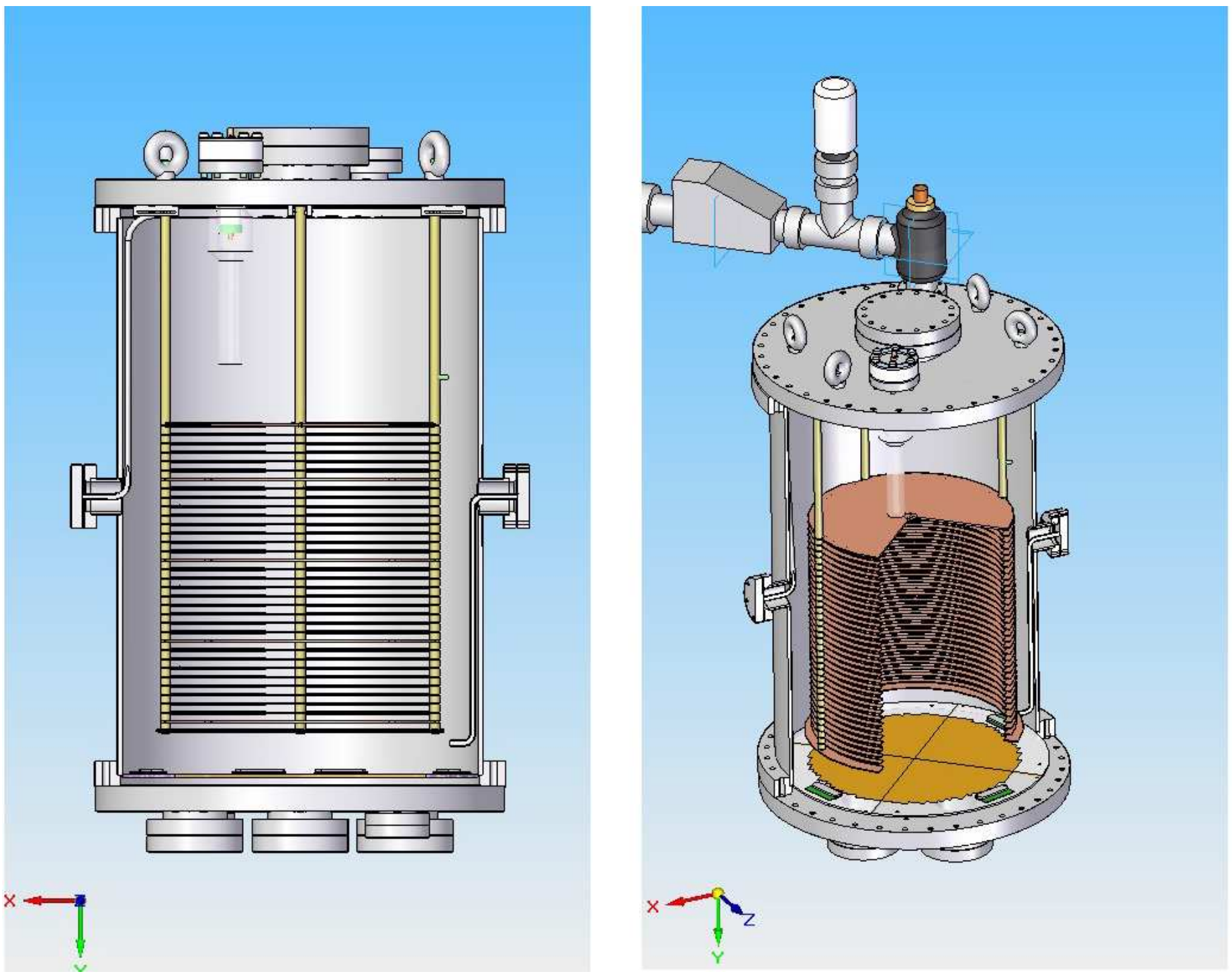

**Figure 1.** Designs of the NEXT-MM pressure vessel and some of its internal components: XY cut view (left) and 3D view (right). Field cage, HV feedthrough and Micromegas plane can be seen inside the vessel, while valves system connecting the prototype to the vacuum system are shown in the upper part

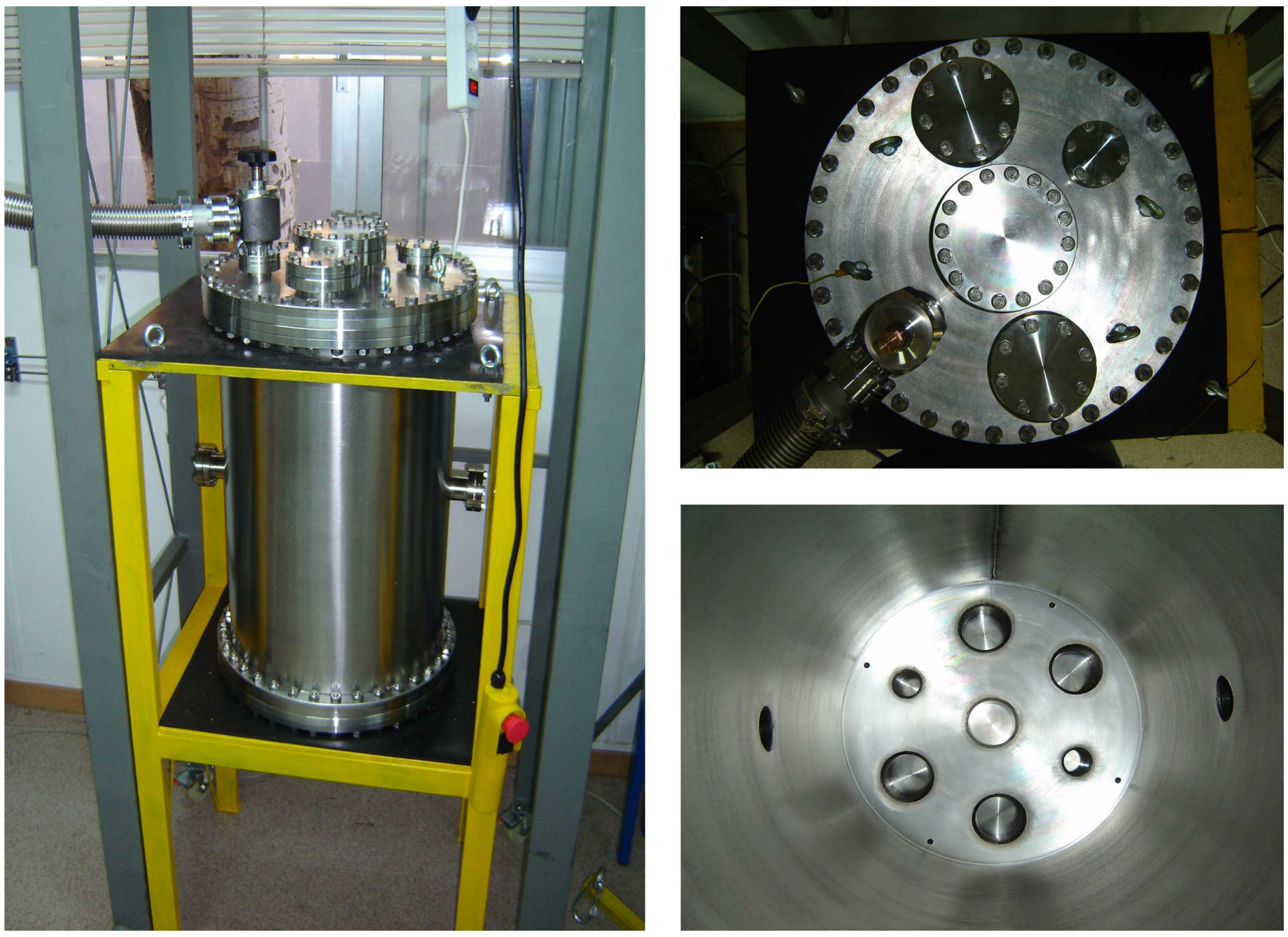

**Figure 2.** General view of the pressure vessel (left), external view of the upper cap (right top) and internal view of the vessel body and lower cap (right bottom).

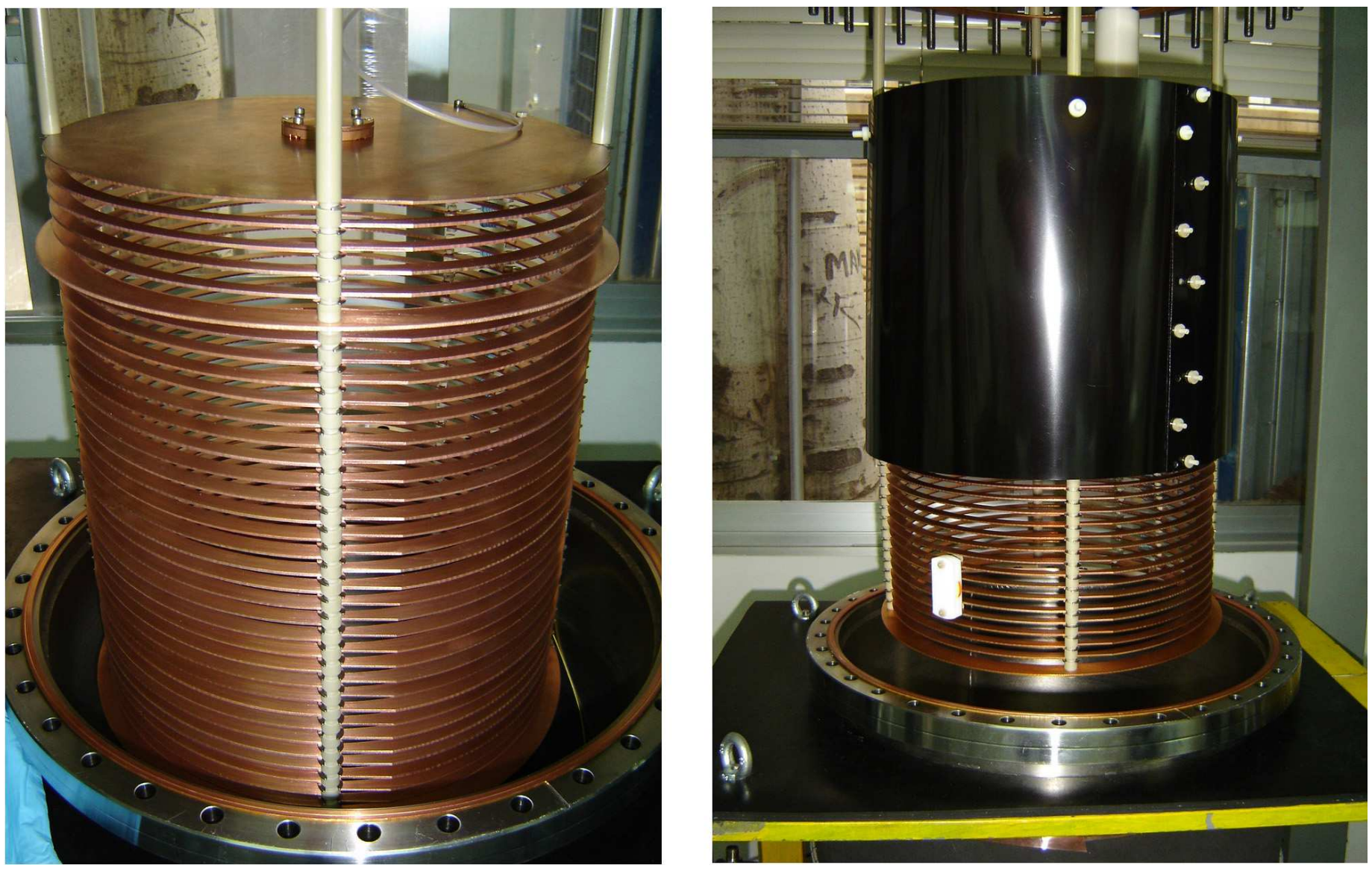

**Figure 3.** Detail of the field cage (left) and the field cage with the Cirlex screen insulator and auxiliary source holder (not used in the presented measurements) installed (right).

### 2.2.2 Readout plane: Micromegas detectors

The data presented in this work were acquired with two different types of Micromegas detectors. For tests presented in Section 3.4, a Bulk Micromegas [16] of 30 cm diameter with 128 $\mu m$ amplification gap was used, covering all the sensitive surface of the TPC (Figure 4-left). The anode of this detector is segmented in 1152 independent pixels of $0.8\times0.8\,\mathrm{cm}^2$ distributed as shown in the design presented in Figure 5. The energy of the event can be determined using the mesh signal, that can be also used to trigger the event, while the study of the signals coming from the different pixels allows the simultaneous reconstruction of the energy of an event and its topology.

The second detector used is based on the Microbulk technology [6]. Four sectorial detectors (with the shape of a quarter of a circle) were installed due to the current manufacturing limitation to a maximum wafer size of $\sim25\times25\,\mathrm{cm}^2$ (Figure 4-right). Each of these detectors, of 50 $\mu m$ gap, 35 $\mu m$ holes diameter and 100 $\mu m$ pitch for the mesh, has a pixelized anode with 288 pixels of $0.8\times0.8\,\mathrm{cm}^2$. It therefore totals 1152 pixels for the whole readout plane as in the Bulk case (Figure 5). Bulk Micromegas provide robust detector, but the fact that Microbulk ones ensure higher field homogeneity and better energy resolution, makes them a more suitable option for NEXT-MM. Additionally, the used raw materials (a simple double-sided copper-clad kapton sheet) and the manufacturing process assure a better radiopurity control of the detector.

### 2.2.3 Other components

1. High voltage cables. Made of a single copper wire as conductor, and Kapton as insulator (which produces a radiopure and low outgassing element), several cables are placed inside the vessel to make the electrical connections between the corresponding feedthroughs and the cathode, the lowest ring of the field cage, and the mesh of the Micromegas detectors. To avoid any soldering inside the vessel (which could increase the outgassing rate inside it and could contaminate the gas), different clamps have been used to make the electric connections.

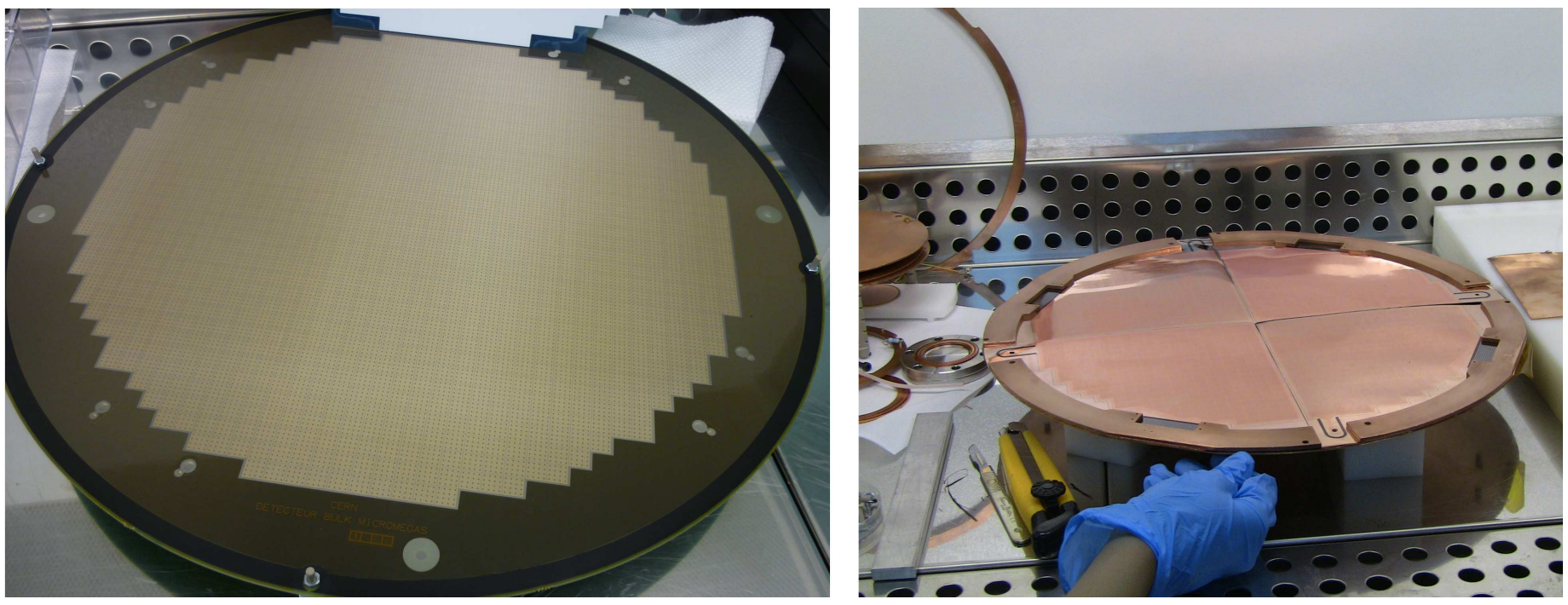

**Figure 4.** Bulk (left) and Microbulk (right) Micromegas used for instrumenting the readout plane of the TPC.

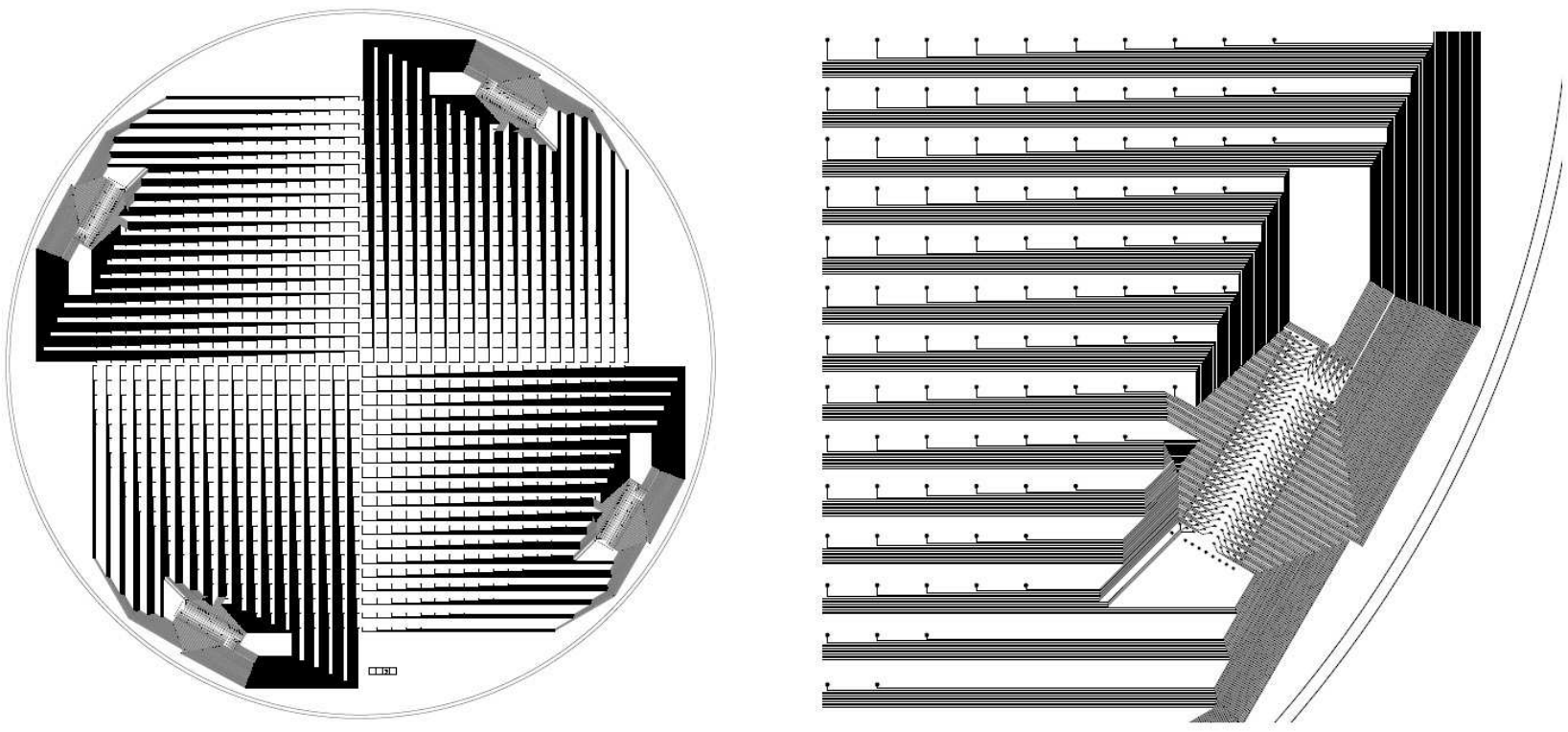

**Figure 5.** Design of the pixels layout and routing for the pixelized anode of the Bulk Micromegas (left) and detail of the routing close to one of the four connectors prepared for the signal readout (right). The routing of the Microbulk Micromegas used and described in the text is equivalent to one quarter of the Bulk design.

2. Pixels readout cables. In order to take out of the vessel the signals coming from the pixels of the Micromegas detectors, flexible flat cables made of copper and Kapton with rigid heads based on FR4 have been used. Each cable allows the routing of up to 300 signals without any soldering, and the connection between the cable and the detector or the signal feedthrough is made by commercial 300-pin solderless connectors (Figure 6-left-top) by SAMTEC (GFZ series) [17]. These cables are also used to route the signal outside the vessel to the data acquisition system (DAQ).

3. High voltage feedthrough. To apply the high voltage to the cathode of the field cage, a feedthrough made of copper and teflon has been developed. Field-cage tests (Section 3.3) show that it can comfortably stand 25 kV (Figure 6-right-top).

4. Mesh signal feedthrough. Commercial 4-SHV connectors feedthroughs have been installed to apply the operating voltage and to extract the signal from the detector mesh, as well as to make any other necessary connections (Figure 6-left-bottom).

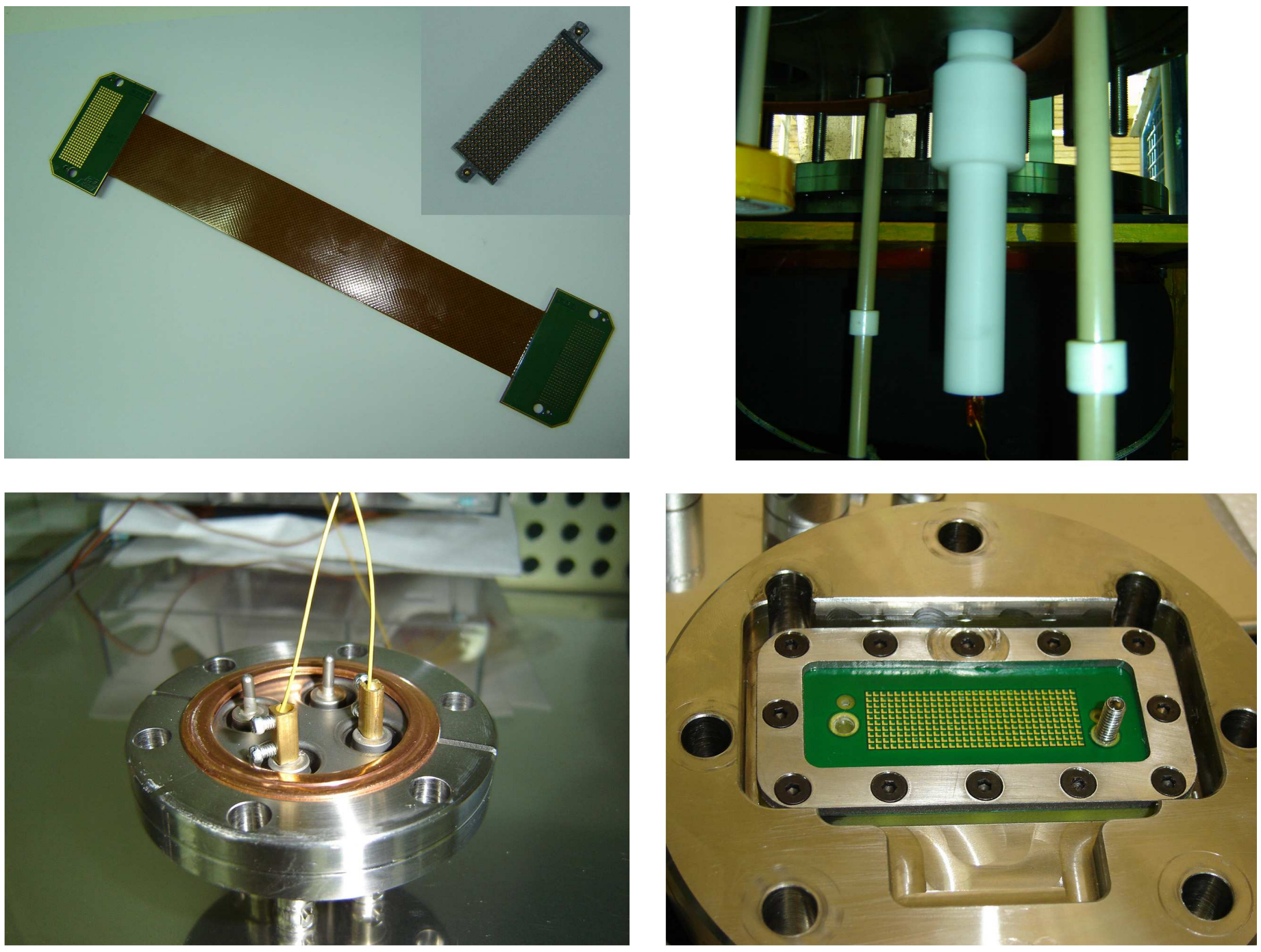

**Figure 6.** Different internal components of the experimental setup: flexible read out cable for pixels signals and solderless connector (left top), feedthrough for the high voltage supply of the cathode (right top), commercial 4-SHV feedthrough for the mesh plane (left bottom) and feedthrough for the pixel signal cables (right bottom).

5. Pixels signal feedthrough. Specially designed to fit the connectors of the flexible cables to them. They have been produced machining a commercial CF port to hold a PCB board with the same profile and footprint as the connector (Figure 6-right-bottom).

6. There are some other small pieces that have been installed to fix the internal components to the vessel based on PEEK, copper Cirlex and Kapton, assuring low outgassing rates

### 2.3 DAQ

The DAQ of NEXT-MM can be divided into two main parts, corresponding to the treatment of the mesh and pixels signals, respectively. For the mesh, the signal is handled in a first stage by a preamplifier 2004 Canberra (PA) which also supplies the voltage to the detector. Signal was subsequently amplified by a 2022 Canberra amplifier (Amp) that provides a measurement of the energy of the event but could also be used to trigger the acquisition of the signals registered at the pixels plane. The pixels are independently digitized using the T2K DAQ, based on the AFTER chip [18].

Figure 7 presents a drawing of the DAQ system. As said, pixel signals are processed through 4 flexible flat cables (based on 300-pixel SAMTEC connectors) that are interfaced to the first stage of the T2K electronics (different front-end concentrators labeled as FEC) with special cable-boards. The FECs are connected to a front-end mezzanine (FEM) that sends the signals to a data concentrator card (DCC),which is connected to the computer where the DAQ software runs and data are stored. The trigger is provided by a logic OR of the amplified mesh signals, after passing a threshold fixed by a low threshold discriminator module (LTD). Additional trigger conditions provided by external detectors are incorporated in some of data taking here reported. The digitalization window width can vary depending on the running conditions (gas pressure and drift field).

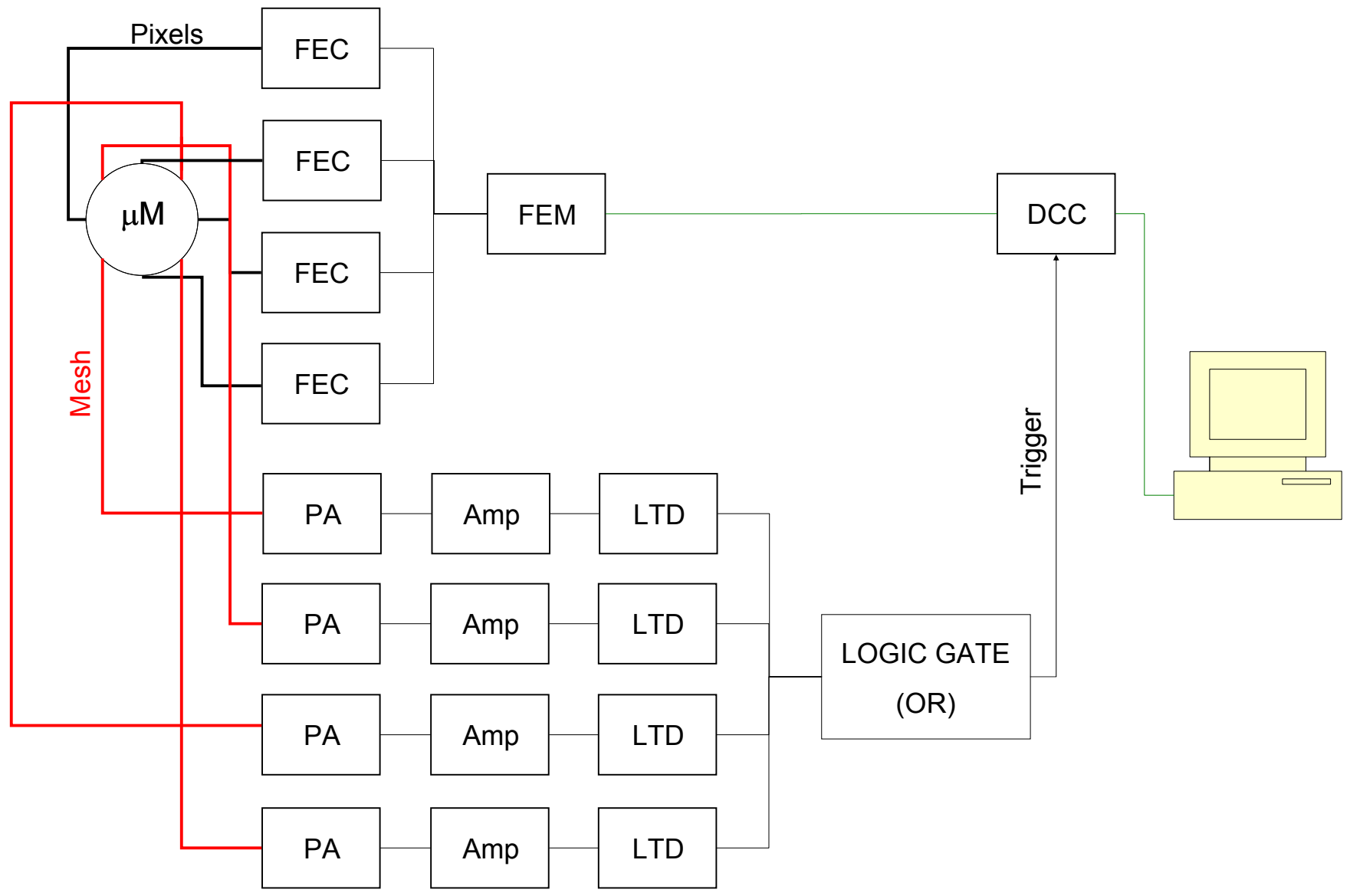

**Figure 7.** Conceptual sketch of the acquisition system of NEXT-MM adapted to four mesh signals as in the case of the Microbulk Micromegas. For the case of the Bulk detector, only one mesh signal has to be taken. If mesh signals have to be registered to obtain an energy spectrum, the amplified signals are processed by a Multi-Channel Analyzer (see text for components description).

### 2.4 Gas and vacuum system

The gas system of the experimental setup has been described in [12]. Its purpose is the recirculation and purification of the gas, and can be divided into several subsystems, as seen in the sketch of Figure 8.

The high pressure system is composed of elements that can work at pressures up to 12 bar and includes the recirculation, purification and recovery subsystems. Prior to any data-taking campaign, this system is pumped down to vacuum. The vacuum subsystem consists on a turbomolecular pump and is isolated from the high pressure part of the system by two consecutive valves, a vacuum valve and an all-metal one. After that procedure, the gas enters the system either from a bottle storage or through a gas mixer, filtered by an Oxisorb purifier.

The system counts with a membrane pump for the recirculation of the gas, which takes the gas from the exit valve of the vessel and forces it through the purification line, before it enters the chamber again. This purification subsystem can use two possible filters, an Oxisorb and a FaciliTorr filter by SAES, but it can be also bypassed when the whole system is being pumped to vacuum. The composition of the gas in the system is periodically checked with a mass spectrometer.

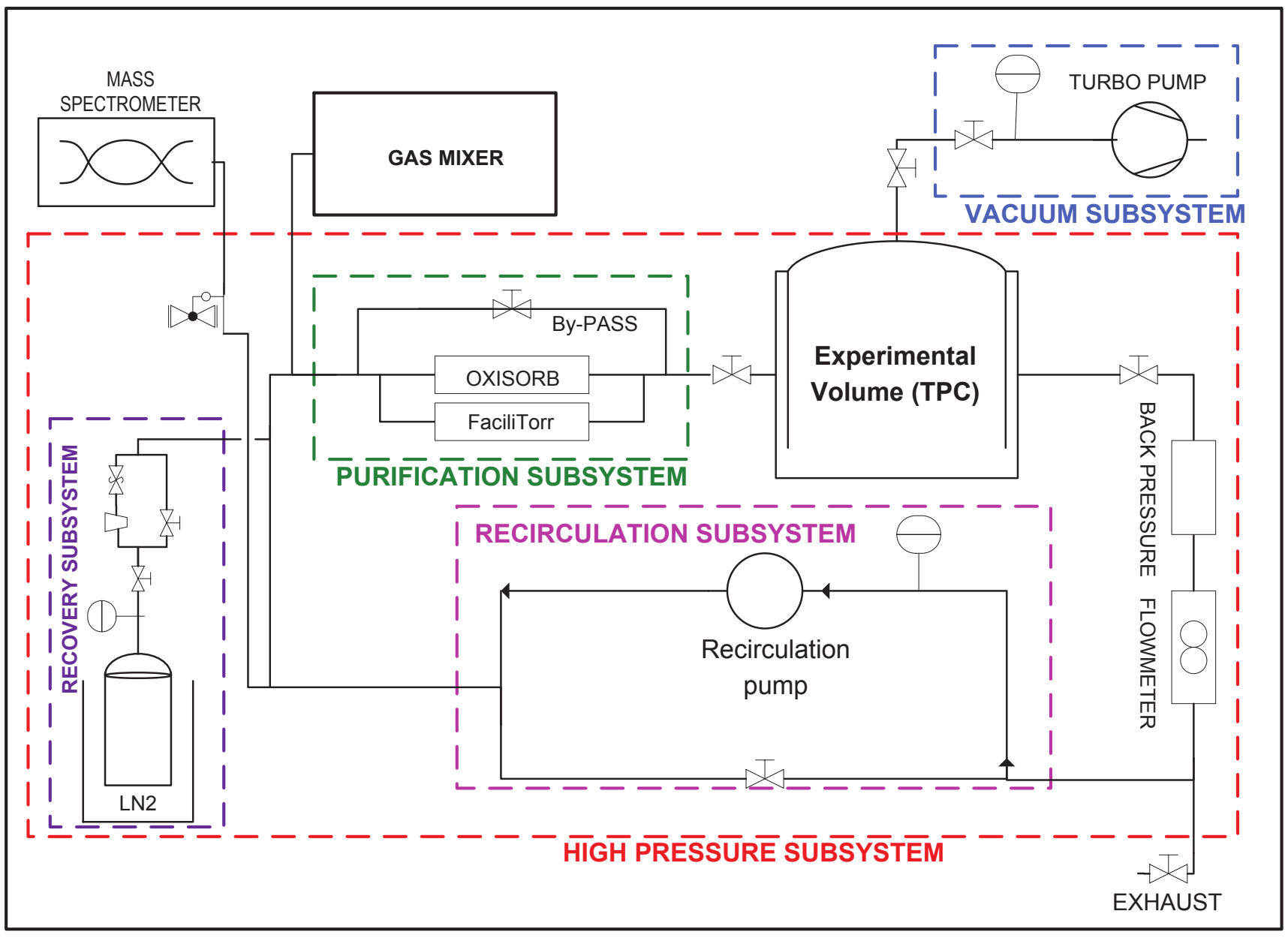

**Figure 8.** Sketch of the gas system used in the measurements, featuring the vacuum and the high pressure subsystems and the different purification, recirculation and recovery components inside the latter one.

At the end of the data-taking, the gas can be reclaimed into a stainless steel bottle, cryo-pumping it using liquid Nitrogen. This recovery branch is also used to insert the gas back into the system. For the data here described a pre-mixed mixture was always used, coming from the recovery subsystem in the case of Xe-TMA.

## 3. Commissioning tests

Before the data taking, several tests were developed in order to check the good performance of the system as well as to safely reach the optimal conditions required to operate the Micromegas detectors regarding gas pressure, outgassing levels and high voltage supply.

### 3.1 High pressure

A pressure test using pure Argon was carried out in order to verify if the vessel itself, but also if the different valves and feedthroughs, and the gas system (described in Section 2.4) can sustain up to 10-15 bar (that correspond to the nominal operation pressure of NEXT).

Figure 9-left shows the evolution of the pressure along 10 days after filling the vessel with 11 bar of pure Argon. The observed oscillations correspond to the variations of the environmental temperature (see Figure 9-left). From a linear fit to the temperature-corrected pressure, $(P/T)\times T_{mean}$,

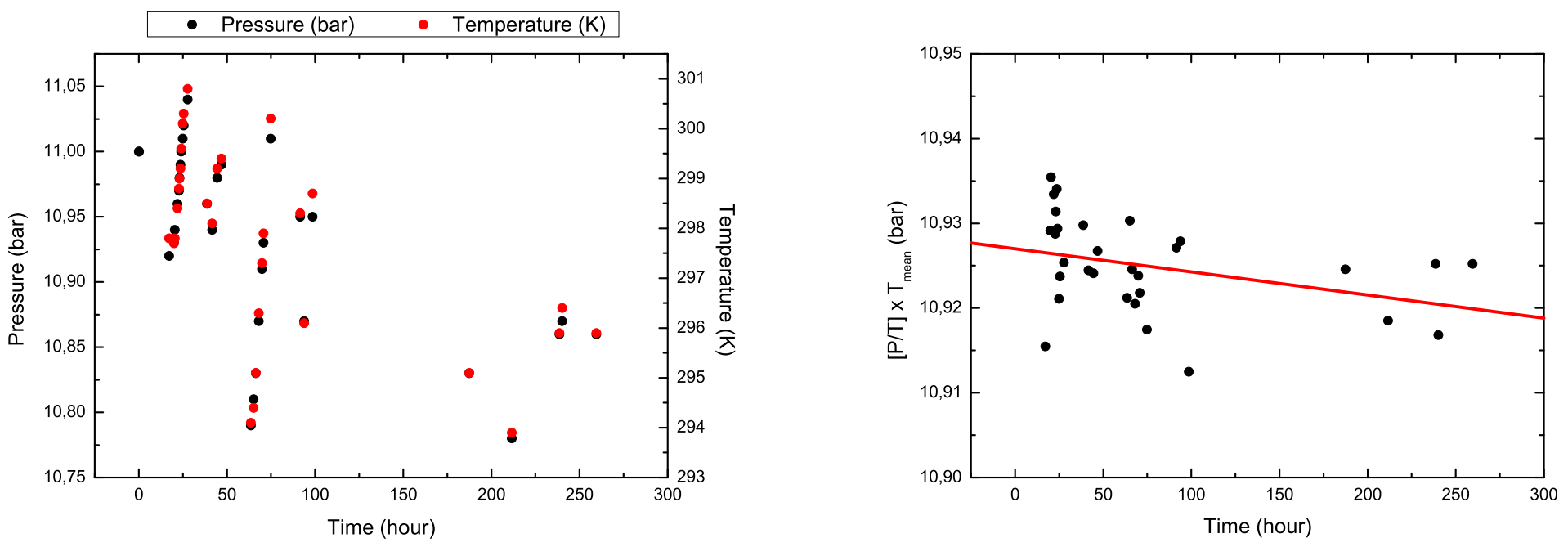

**Figure 9.** Time evolution of the monitored temperature and pressure of NEXT-MM during the high pressure test (left), and evolution of the ratio P/T normalized by the mean temperature of the test. The measured leak rate is below $5.9 \times 10^{-4}$ mbar×l×s$^{-1}$, limited by the accuracy of the high pressure manometer (right).

system leaks are estimated to be smaller than $5.9 \times 10^{-4}$ mbar×l×s$^{-1}$, that satisfies the operation requirements.

### 3.2 Vacuum and outgassing

The filling of the vessel with gas must be performed in good vacuum conditions. In addition, the release of trapped air or other impurities from the internal materials could also degrade the quality of the gas, so it is necessary to minimize the outgassing rate of the whole system. One way to improve sequentially this outgassing rate is the realization of bake-out cycles to facilitate the release of impurities while the system is pumped. In order to evaluate the contribution of each internal element to the total outgassing rate, a bake-out cycle was performed after the installation of a component. Elements with higher thermal resistance were installed first allowing the heating of the system at higher temperatures.

To get a reference vacuum value, initially, the system was pumped without any bake-out cycle. After 95 hours of pumping, a pressure of $7.8 \times 10^{-7}$ mbar was reached, which is in good agreement with the specifications provided by the manufacturer.

The first bake-out cycle was performed with the vessel empty. The vessel was heated up to 180 °C while it was pumped for almost 140 hours. After that, the heating system was switched off and the pumping continued till the vessel reached the environmental temperature. The pressure obtained was $6.3 \times 10^{-7}$ mbar, improving the reference value. After the pumping system was switched off, the outgassing was measured to be $4.7 \times 10^{-7}$ mbar×l×s$^{-1}$.

Table 1 summarizes the parameters of the bake-out process done (including the first one already described): the time that the system is simultaneously heated and pumped (heating time), the heating temperature, and the time the system is pumped after the heating system is switched off (pumping time) giving an initial pressure value of the system labeled as $P_0$.

The obtained results do not reveal the existence of any component with an abnormal high outgassing rate, showing that when longer the bake-out time is for the common components (like the vessel or the field cage) better outgassing rates are obtained. Therefore, regular bake-out cycles could be necessary to keep or even improve the outgassing rates reached in the prototype.

**Table 1.** Summary of the bake-out cycles carried out for NEXT-MM (indicating the components inside the vessel, see text for a description of the full setup) and the corresponding outgassing rate measured for each case. Errors in the measurement of $P_0$ and outgassing are negligible.

| Components | Heating Time (h) | Heating Temp. (°C) | Pumping Time (h) | $P_0$ (mbar) | Outgassing (mbar × l× $s^{-1}$) |
|---|---|---|---|---|---|
| Vessel | 140 | 180 | 26 | $7.4\times10^{-7}$ | $4.7\times10^{-7}$ |
| Field Cage | 100 | 155 | 17 | $3.8\times10^{-7}$ | $3.4\times10^{-7}$ |
| F. Cage + Resistors | 113 | 140 | 6 | $9.8\times10^{-8}$ | $5.9\times10^{-8}$ |
| F.C + Rs + Bulk $\mu$M | 94 | 150 | 144 | $6.1\times10^{-8}$ | $3.2\times10^{-8}$ |
| Full setup | 112 | 150 | 11 | $7.6\times10^{-7}$ | $4.5\times10^{-7}$ |

The outgassing rate of the full setup (composed by the detector, the field cage with the resistors and the Cirlex protection, all the necessary internal cables and the required feedthroughs described in Section 2.2 was) $4.5\times10^{-7}$ mbar×l×$s^{-1}$, enough to keep the contamination of operation gas under acceptable levels.

### 3.3 High voltage

Micromegas HV is supplied by a commercial SHV feedthrough, allowing to supply up to 3.5 kV, being well above the typical operational voltage for Micromegas (few hundreds of volts). The field cage cathode could be eventually biased up to 35 kV to have electric fields of the order of 1 kV $cm^{-1}$ with gas pressures that could go up to 10 bars.

The electric current passing through the field cage was monitored while the high voltage was increased. The breakdown point is defined as the voltage where the current begins to be higher than the expected linear behavior (based on the total resistance of the field cage) which indicates that some current leak to ground (defined by the vessel) is taking place.

Figure 10-left shows the evolution of the current, while Figure 10-right summarizes the evolution of the breakdown point for different pressures. For pure Argon an operating pressure of ∼12 bar is needed to reach the 35 kV, while ∼32 kV could be reached for 10 bar. Assuming that the breakdown value for Xenon is approximately 2 times higher than for Argon, Xenon could hold fields greater than 1 kV $cm^{-1}$ for pressures above ∼4.5 bar. The extrapolation of this data to Xenon mixtures, as the one used in Section 4, is not so straight forward and studies are ongoing to measure the breakdown point for different pressures for Xe-TMA mixtures.

### 3.4 DAQ and readout

Based on previous experience [12], in order to perform a fast and reliable test of the DAQ system and the readout planes, a first data taking was devised with 1 bar of premixed Ar-(2%)$iC_4H_{10}$ mixture.

For the Bulk Micromegas described in Section 2.2.2, $^{222}$Rn coming from a $^{226}$Ra source was diffused in the gas, registering the amplified mesh signal with a Multi-Channel analyzer to obtain the energy spectrum at different drift field conditions. Figure 11-left shows the three expected alpha emissions of the $^{222}$Rn decay, with energies of 5.5, 6.0 and 7.7 MeV, together with a tail at lower energies, corresponding to not fully contained events in the sensitive volume. The evolution of the 5.5 MeV peak area (fitting it to a Gaussian distribution over a flat background), is compatible with an exponential decay with the half live of the $^{222}$Rn as showed in Figure 11-right.

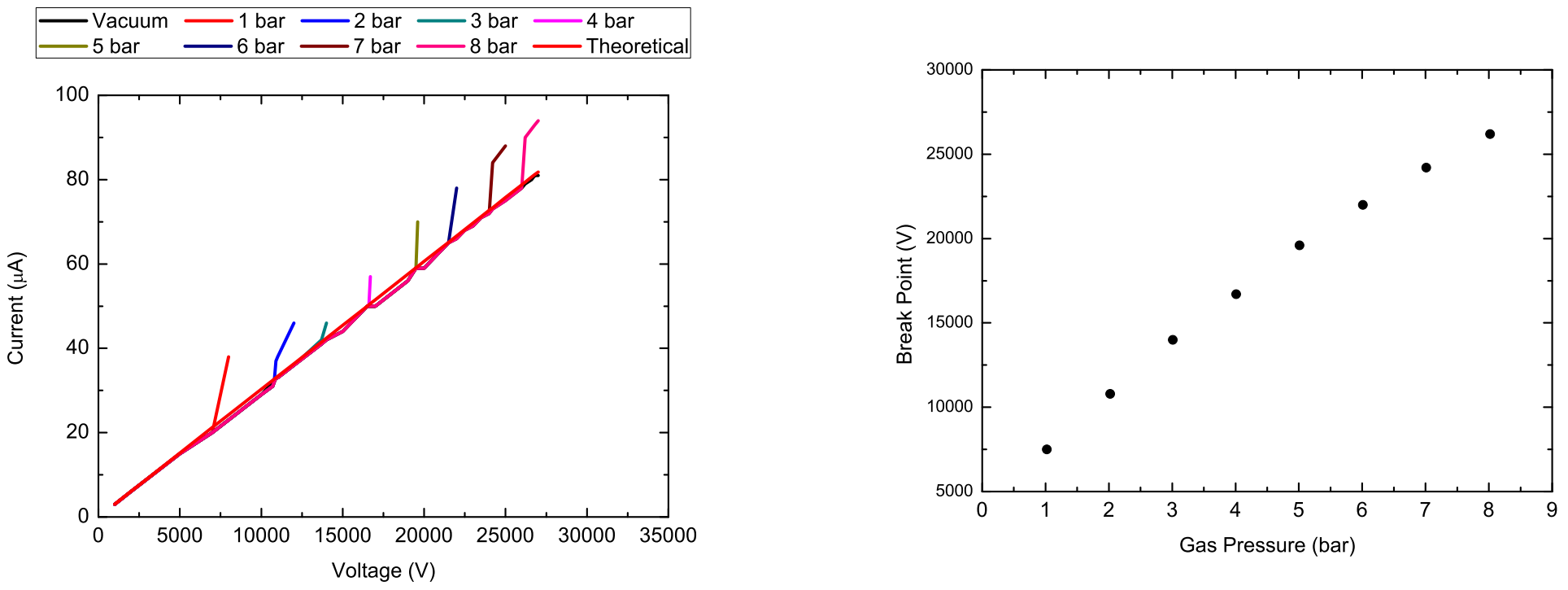

**Figure 10.** Current and breakdown point evolution depending on the system pressure.

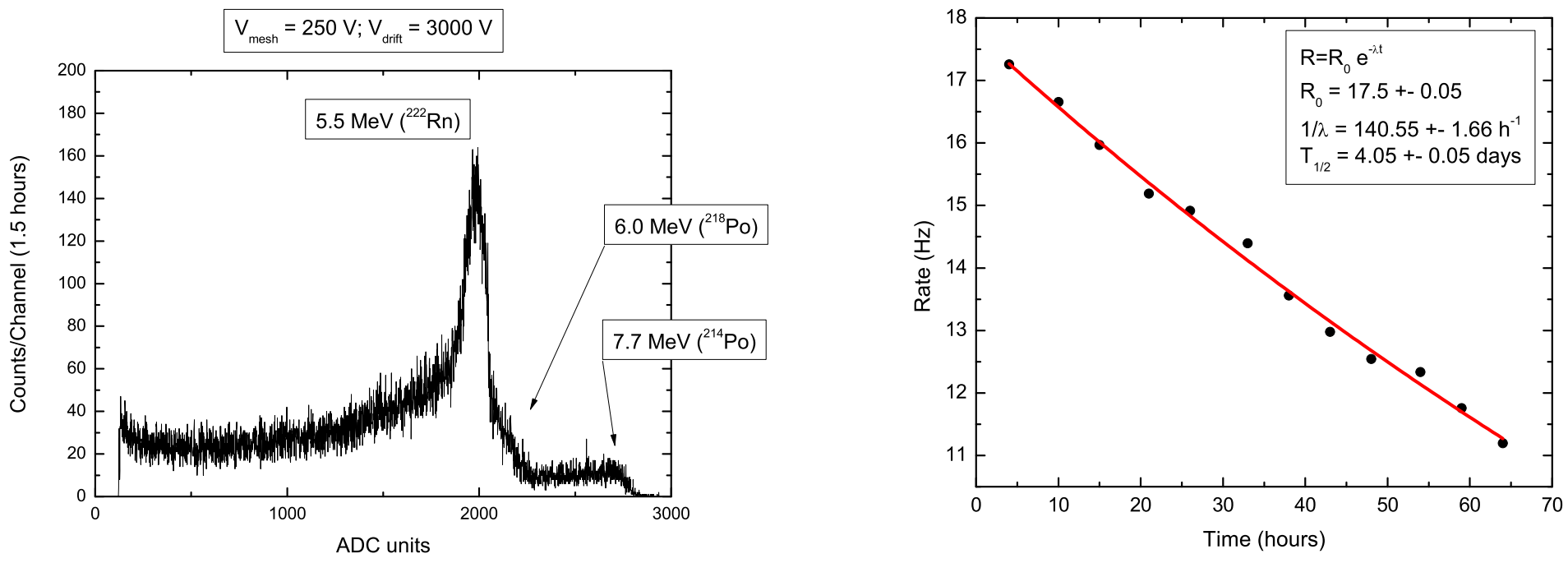

**Figure 11.** $^{222}$Rn energy spectrum obtained from the mesh signal of a Bulk Micromegas operated at 1 bar of premixed Ar-(2%)i$C_4H_{10}$ mixture (left). Time evolution of the detection rate of the $^{222}$Rn 5.5 MeV alpha (right).

Different data takings have been carried out keeping constant the amplification field and varying the drift field from 85 to 228 V/cm/bar obtaining spectra equivalent to the one shown in Figure 11-left. The total gain of the TPC depending on the drift field was estimated from the position of the 5.5 MeV emission peak. Variations of the gain will indicate the presence of attachment or recombinations effects in the gas. Observed variation is at the few %-level for Ar-(2%)i$C_4H_{10}$ at 1 bar and in the range 80-240 V/cm.

Together with the outgassing values presented in Section 3.2, this test reveals that the level of impurities inside the sensitive volume is low enough to have a good quality of data. Further studies using different Xenon mixtures and high pressures are ongoing.

A second commissioning test was carried out with the Microbulk detectors presented in Section 2.2.2. An $^{241}$Am was placed at 8 cm distance from the center of each readout quadrant. The position of the sources, with an alpha particle of 5.5 MeV as main emission, together with the use of Ar-(2%)i$C_4H_{10}$ gas, assure the detection of fully contained events for all the detectors with low values of drift and amplification fields.

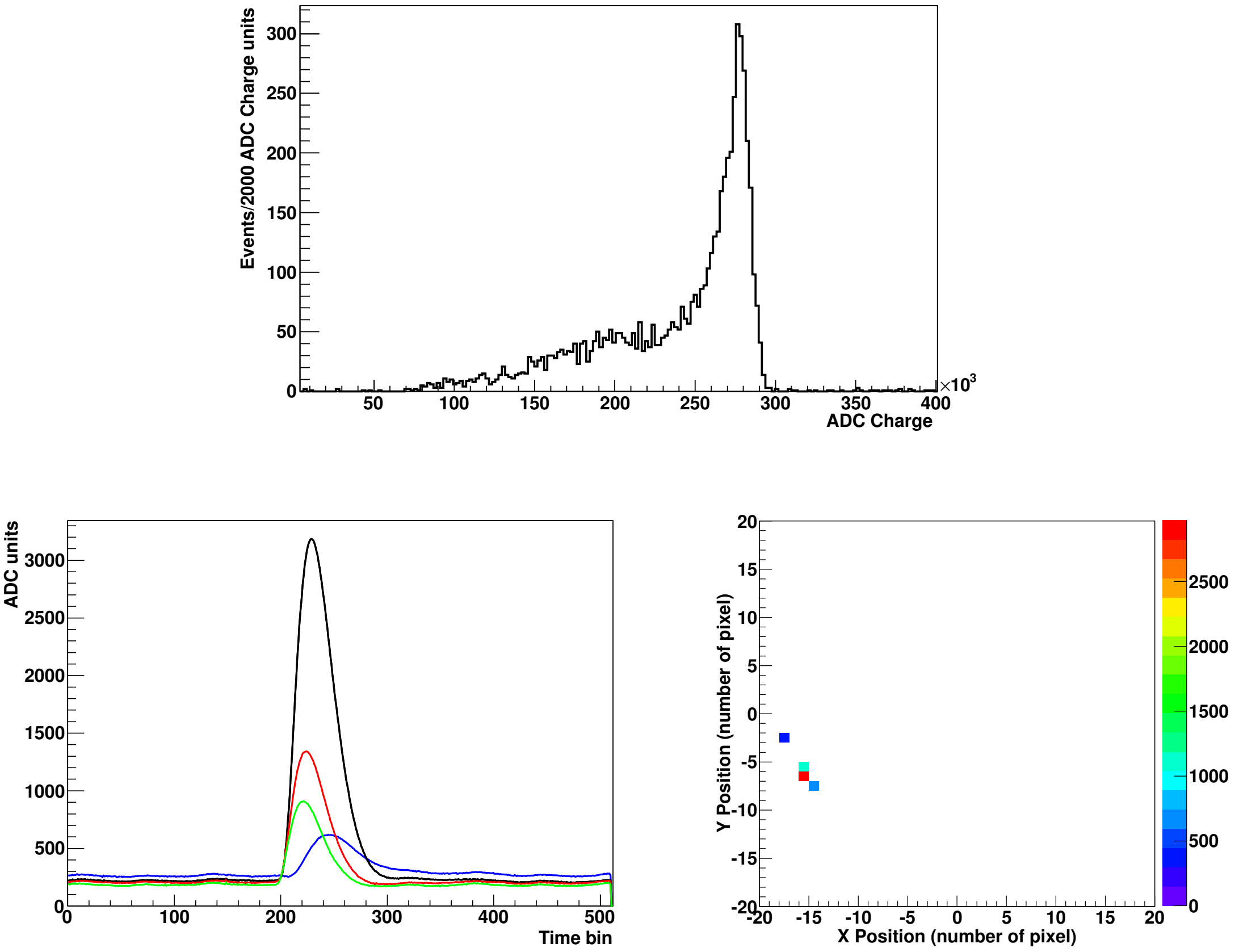

**Figure 12.** $^{241}$Am energy spectrum obtained from the pixel signals of Microbulk Micromegas operated at 1 bar of a premixed Ar-(2%)i$C_4H_{10}$ mixture (top) and example of a ∼5.5 MeV event: pulses in the activated pixels (bottom left) and 2D projection of the track in the XY plane, indicating (in ADC charge units) the total charge of the pixel with the color scale (bottom right).

The mesh signal was used as trigger and the energy spectra were obtained by adding the charge of the pixels that passed a certain threshold (by integrating each pulse over the digitization window). The energy spectrum obtained at 1 bar for Ar-(2%)i$C_4H_{10}$, is presented in Figure 12 for one of the tested detectors as well as an example of pixel pulses and the tracks reconstructed of a 5.5 MeV alpha event. The spectrum shows the main peak of the 5.5 MeV emission with a continuous tail at lower charge (lower energy) corresponding to alpha particles that have lost some energy before interacting in the sensitive volume or that have not been fully contained in the sensitive surface of the Micromegas detectors. The obtained energy resolution is compatible with the expected value for these detectors operated [19], nevertheless a small degradation was observed due to the connectivity of the pixels lines to the DAQ. The loss of some of these signals prevented the full detection of the energy for some events.

A dedicated study of the pixel connectivity was performed. The level of connectivity achieved in the feedthrough is nearly 100 %, but the tighter mechanical constraints imposed by the very thin 50 $\mu$m-thick copper-clad Microbulk Micromegas foils caused misalignments between the footprint and the connector, together with loss of parallelism . Systematic capacitive measurements performed with an LCR-meter in the 1 kHz range, provided a map the low-connectivity regions, allowing the detection of the problem. After some *ad hoc* solutions, the level of connectivity at the Micromegas footprint reached 95 %.

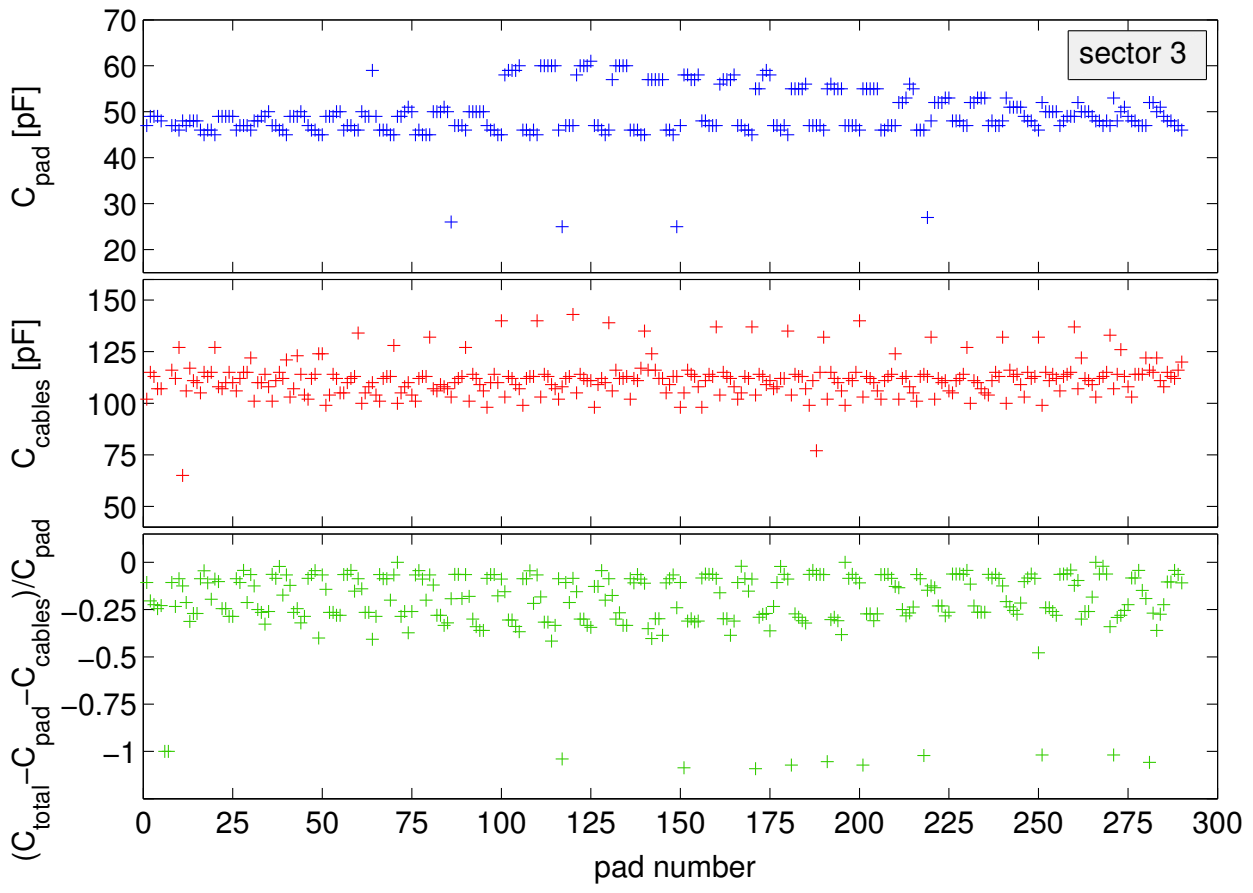

**Figure 13.** Measured capacitance with respect to ground for each separate pad (up). Measured capacitance for the cables up to the front-end electronics (middle). Difference between the measured and expected capacitance, normalized to the pad capacitance. When the connection to the pad is missing, a clear $-1$ is visible (down). The level of connectivity achieved for the sector in figure is above 95 %.

Figure 13 illustrates the experimental procedure of the capacitive measurements for one of the sectors. The capacitance was measured with respect to the ground line, yielding a 'pad capacitance' in the 50-60 pF range (Figure 13-up), that includes the expected contribution of the pad capacitance to the mesh plane ($\sim$30 pF, highly coupled to ground), the parasitic coupling to the support copper plate ($\sim$25 pF, separated by additional 100 $\mu$m Kapton foil) with deviations from this number stemming from different trace routing. Four pads are halved by design, and they can be clearly identified in Figure 13-up. The remainder of the signal line down to the front-end electronics includes the flat cables and the feedthrough, and showed some dispersion in its coupling to ground, being centered around 110 pF (Figure 13-middle). Overall, the two contributions (flat cable + pad) represent fairly faithfully the total capacitance of the signal line, and hence a bad connection could be clearly identified after assembly as that showing a value of $-1$ in the representation of Figure 13-down, indicating a 'missing pad'.

## 4. First results with Xe-(2%)TMA

To study the response of the Microbulk Micromegas to Xe-(2%)TMA gas mixture, the detectors were installed inside the vessel with Xe-(2%)TMA at 1 bar recirculated through a SAES purifier at a rate of 12 ln/h. This mixture presents a lower diffusion than pure Xenon. It has been proven that the addition of small fractions of TMA in Xenon increases the gain of the detector and helps to improve the energy resolution at high pressures [12]. It could also reduce the Fano factor in Xenon-based chambers, as suggested in [9].

An $^{241}$Am calibration source was placed at the center of the cathode (see Figure 3-left). The source was coupled to a Si detector to detect the alpha emission that will be used as trigger. 59.5 keV gamma events that accompanies one third of the times the emission of the alpha particle, were triggered with the coincidence of a signal in the Si sensor and the mesh. The Si also provides the strat time of the event ($t_0$). The source was placed at 38 cm from the Microbulk detectors (3 out of 4 where instrumented for these measurements).

The operation voltage of all the Micromegas was 270 V (corresponding to an amplification field of ∼54 kV/cm). The results here shown were taken at a drift field of 150 V/cm.

The energy and the tracks of each event are determined from the charge collected on the pixels, which simplifies the DAQ in terms of synchronization to assign to one event the signals coming form the mesh and the pixels (independently obtained). An example of a background electron which crosses three detectors can be seen in Figure 14. Pulses from the electronics are shown on the left, while the 2D projection of the event on the Micromegas plane is shown on the right. The 3D reconstruction shown in the lower graph is done with the relative time of each pixel. After tuning and measuring the delays introduced in the trigger chain, the time distribution of the events allows the identification of the drift time from the cathode to the anode plane between 40 $\mu$s and 150 $\mu$s (Figure 15).

The energy spectrum is obtained adding the charge of all the recorded pixels that pass a 1 keV threshold (based on previous calibrations), obtained from the pulse integral. In Figure 16, the 59.5 keV peak of the $^{241}$Am gammas can be seen along with the ∼29 keV escape peaks corresponding to the Xenon characteristic X-rays ($K_\alpha$ at 29.7 keV and $K_\beta$ at 33.6 keV). These escape peaks overlap with the other main gamma emission of the source at 26.4 keV. The selection criteria of the events are: a) no saturated events; b) events with multiplicity (number of pixels fired) higher than 12 are rejected as they would not belong to the source gammas, and c) events only within the expected drift region (between 40 $\mu$s and 150 $\mu$s based on the observation presented in Figure 15) are considered. In order to avoid border effects, an area smaller than the sensitive one, corresponding to the detector surface, is considered. Two regions of 10×9 and 7×5 pixels respectively have been selected (Figure 18). After applying the described selection criteria in the events confined in these two regions, the spectra shown in Figure 17-left for the 10×9 pixels region, and in Figure 17-right for the 7×5 pixels region are obtained.

The energy resolution for the Micromegas region of 10×9 pixels (that corresponds to 32% of the detector area) is 13.9% FWHM for the 29 keV peak and 11.7% for 59.5 keV. Notice that the 26.4 keV of the $^{241}$Am peak superimposed on the $K_\beta$ escape peak starts to be discernible over the main $K_\alpha$ escape. For the smaller region of 7×5 pixels (12 % of the detector surface), the results improve to 11.6 % FWHM and 9.9 % FWHM for the 29 and 59.5 keV peaks respectively. This behaviour indicates the existence of gain differences between pixels. An inter-pixel gain calibration is ongoing.

Figure 19 shows, for illustration, events of 29 and 59.5 keV respectively for one of the installed Micromegas. The performance of the other detectors is similar.

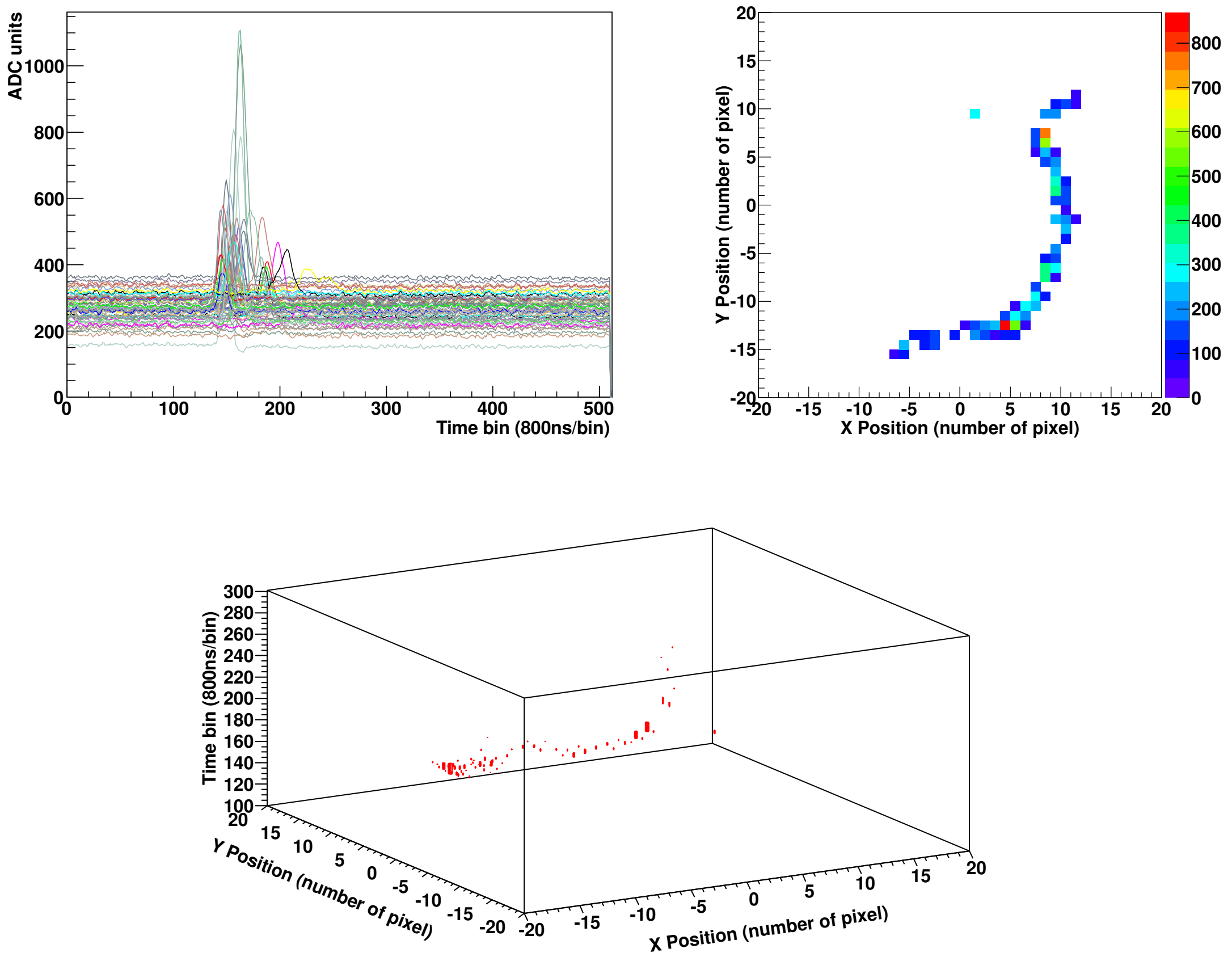

**Figure 14.** Example of background event: pulses in the activated pixels (left), 2D projection of the track in the XY plane, indicating (in ADC charge units) the total charge of the pixel with the color scale (right), and a 3D reconstruction of the event with a relative z position (bottom).

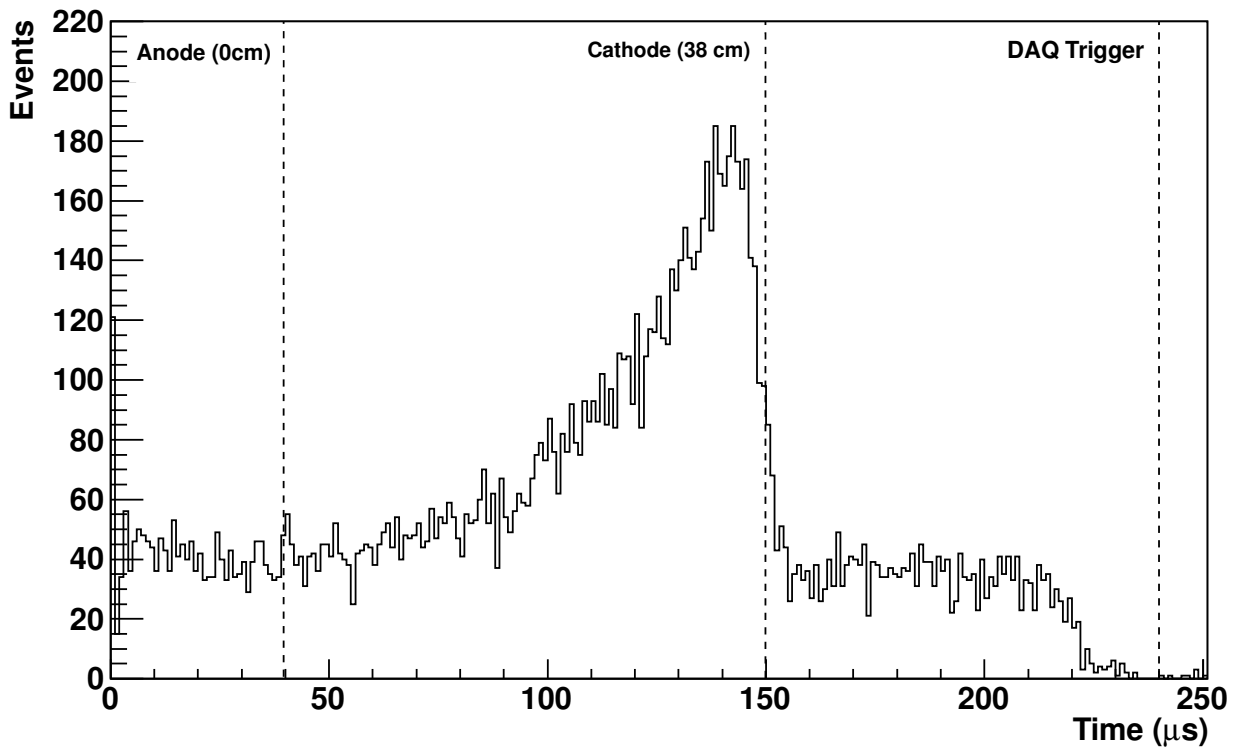

**Figure 15.** Time distribution of the events registered in one Micromegas. The anode plane starts at 40 $\mu$s, according to the trigger delay, and the cathode extends up to 150 $\mu$s, where the maximum of the distribution is situated.

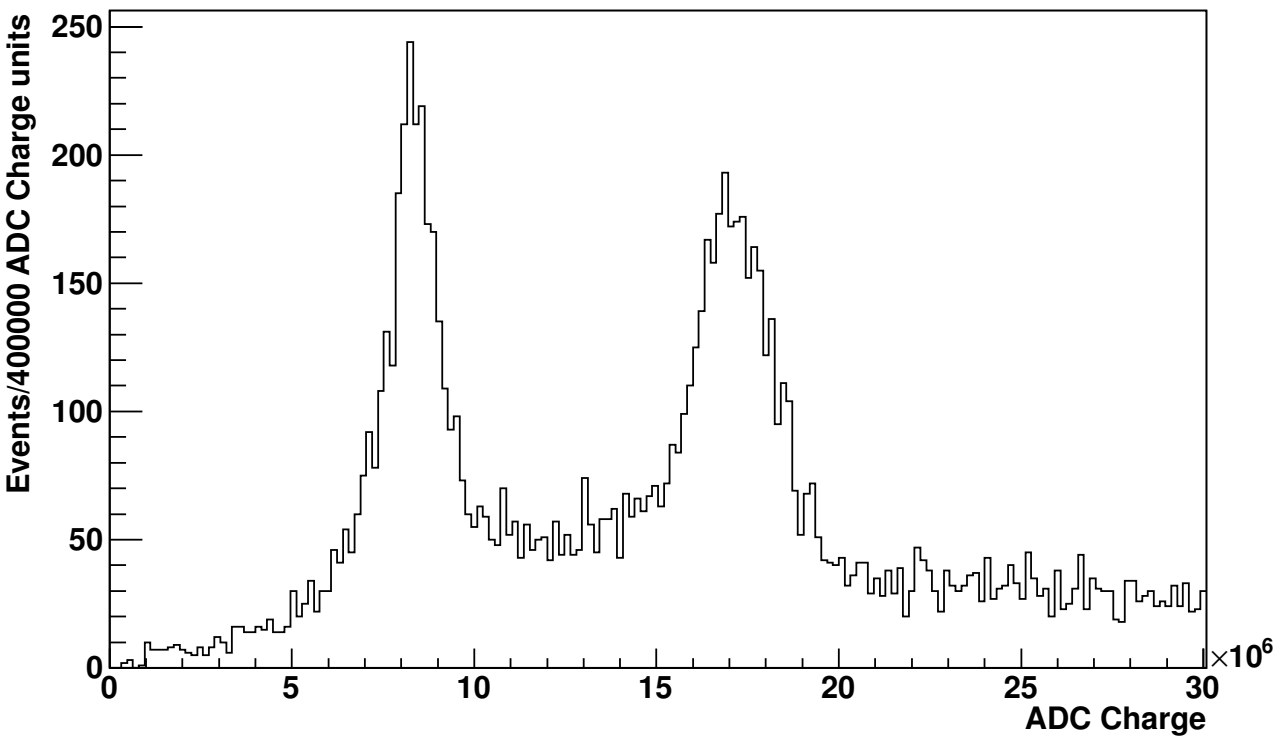

**Figure 16.** Raw spectrum (before applying any selection criteria) registered by one Micromegas in NEXT-MM where the peak of the $^{241}$Am 59.5 keV gamma can be seen. The escape peaks corresponding to the characteristic X-rays of Xenon are also seen around 29 keV.

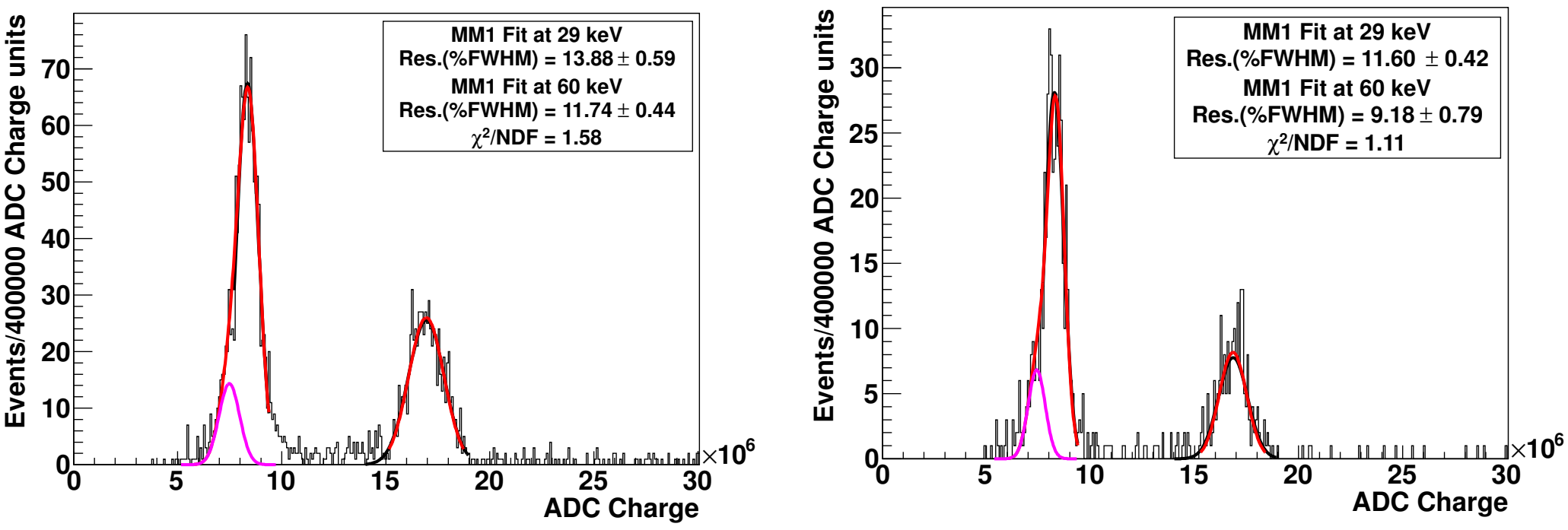

**Figure 17.** Energy spectra after the application of the selection criteria for two different areas of the Micromegas. Left: an area of 10×9 pixels corresponding to ∼32 % of the detector surface was selected. An energy resolution of 11.7 % FWHM is obtained for the 59.5 keV peak. At ∼29 keV the peak is a sum of the escape peaks from the characteristic X-rays of Xenon and the $^{241}$Am gamma at 26.4 keV. Two gaussian fits are used and the energy resolution obtained for 29 keV is 13.9 % FWHM. Right: when a more restricted region is studied (7×5 pixels, i.e., 12 % of the sector's surface), the obtained resolution is 11.6 % and 9.9 % FWHM for 29 and 59.5 keV respectively. This indicates that inter-pixel gain variations are important.

## 5. Outlook and prospects

The results obtained up to date demonstrate that NEXT-MM is ready to perform measurements with different gas conditions using Micromegas detectors registering simultaneously the energy and the topology of the event. The Microbulk technology is expected to show better detection features than the Bulk one [6].

First data-taking has helped to understand the different detector properties and their behaviour. Recirculation of the gas through a SAES purifier or testing of the pixels connectivity have proven to be very useful to improve the data quality. Further plans include data taken with Xe-TMA mixtures at different pressures up to 10 bars. Study of different parameters of the Xe-TMA mixture, like the diffusion coefficient and the drift velocity, together with their evolution with the pressure, will be carried out. Campaigns with other mixtures and pure Xenon are foreseen.

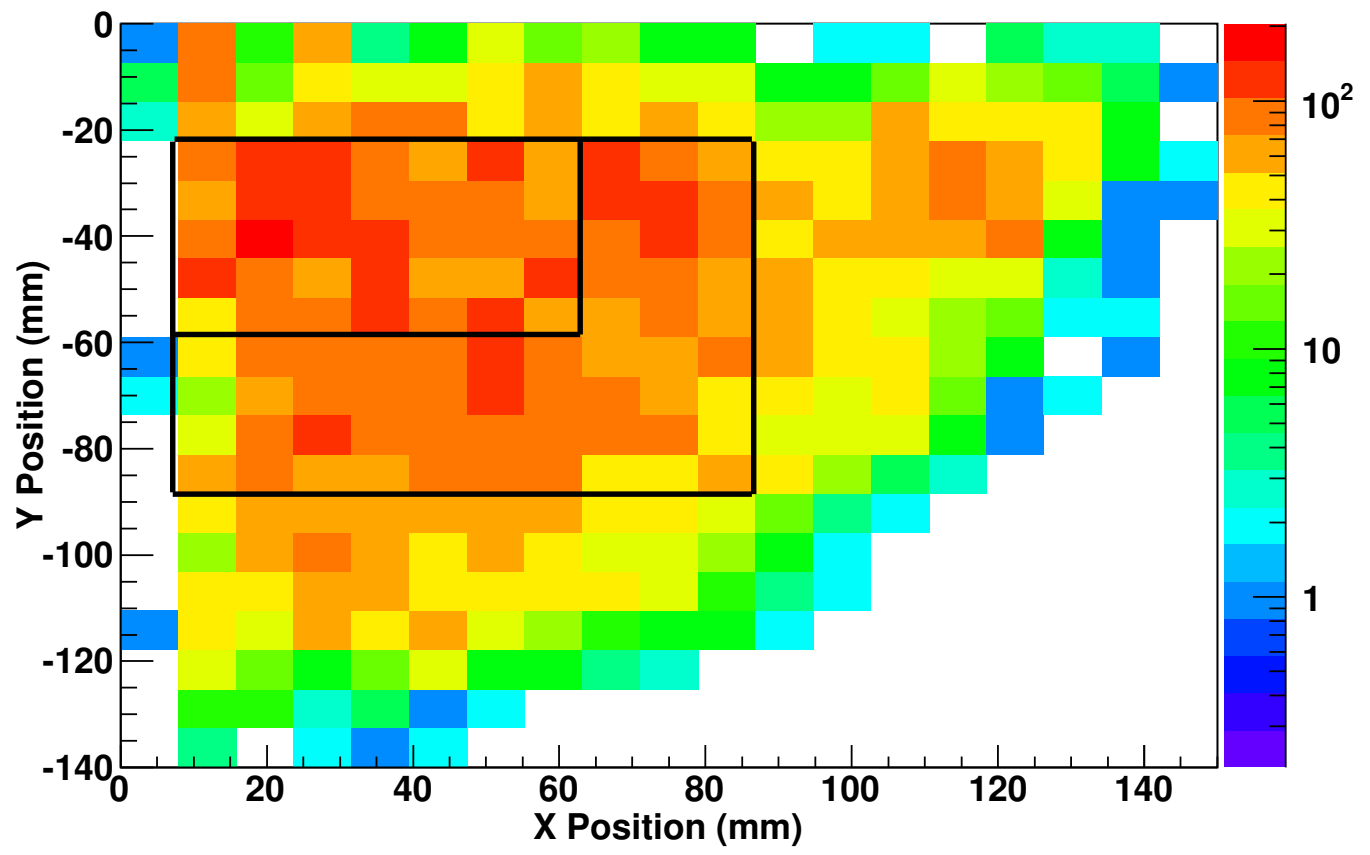

**Figure 18.** Map of the fired pixels in one of the Micromegas detectors installed in NEXT-MM (color of the pixel corresponds to the number of times it has been fired). The two regions selected to obtain the energy spectra after the analysis (of 10×9 and 7×5 pixels, respectively) are indicated.

In parallel to the detector operation and routine data-taking, a deeper development of the analysis tools will be carried out. The first results presented here are promising, but pixel-to-pixel calibration for the Micromegas detectors will improve the energy resolution

## 6. Conclusions

This work presents a detailed technical description, commissioning and first data taken with a general purpose TPC instrumented with Micromegas technology (NEXT-MM). Its readout plane is made by a sectorial arrangement of 4 single wafers manufactured with the Microbulk technique. It contains a total of 1152 pixels in a $0.8 \times 0.8\,\mathrm{cm}^2$ square pattern, that are read out with standard TPC electronics based on the AFTER chip. NEXT-MM is currently envisaged for the study of optimized Xenon-based gas mixtures for $0\nu\beta\beta$ searches.

Functional tests have been summarized, validating the specifications of the detector in terms of resistance to vacuum and high pressure conditions as well as the associated bake-out cycles, high voltage insulation, leak tightness, outgassing rates and overall gas quality. The prototype implements some original technical solutions intrinsically radiopure apart from the Micromegas detectors themselves; most notably: a coaxial copper rod/teflon HV-feedthrough (tested up to 25 kV), as well as the high-density solder-less signal feedthroughs.

The commissioning runs included a readout based both on Bulk and Microbulk Micromegas, with the TPC operated in a standard Ar-(2%)$iC_4H_{10}$ benchmark mixture. Subsequently, first data with Microbulk Micromegas in Xe-(2%)TMA have been presented. Data include long background electron tracks and alphas from a $^{222}$Rn source, illustrating the tracking capabilities of the system. Low energy electron tracks from a $^{241}$Am source (26 keV, 59.5 keV) were reconstructed at 1 bar with a raw energy resolution (FWHM) around 11.6 % at 29 keV.

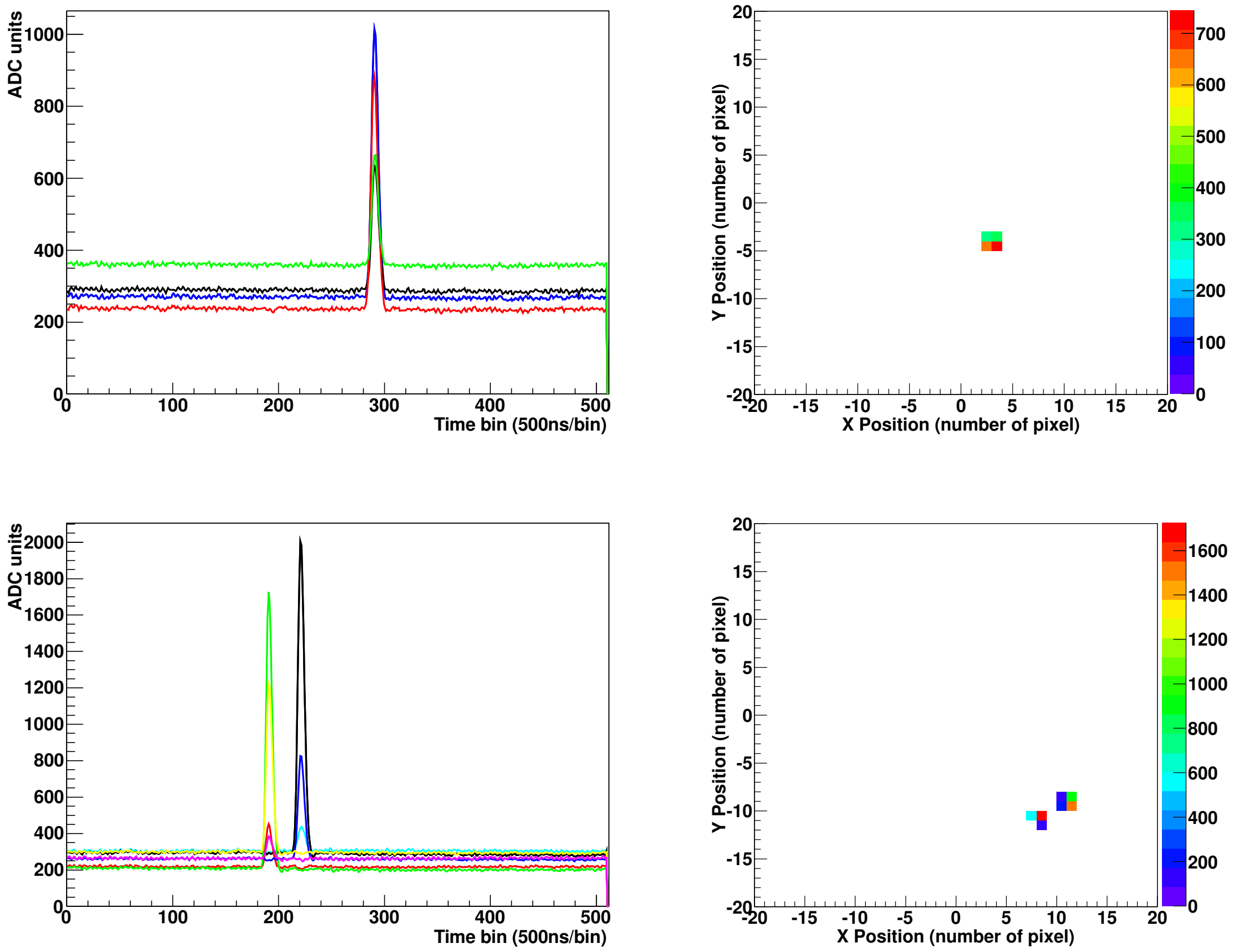

**Figure 19.** Examples of events attributed to the $^{241}$Am source, in the upper part a 29 keV event, in the lower part a 59.5 keV event with two clusters. Pulses in the activated pixels are shown on the left, while the 2D projections of the tracks in the XY plane, indicating (in ADC charge units) the total charge of the pixel with the color scale, are shown on the right.

In summary, NEXT-MM has been successfully commissioned, and first results corroborate the stability of operation of Micromegas-read with Xe-(2%)TMA at 1 bar. Work with the detector is actively going on. Next steps include to carry out longer runs, taking data at higher pressures and using a higher energy photon peak, study event topologies, test different gas conditions, and extract information on gas parameters (drift velocity, diffusion coefficients, etc).

## Acknowledgments

The NEXT Collaboration acknowledges funding support from the following agencies and institutions: the Spanish Ministerio de Economía y Competitividad under grants CONSOLIDER-Ingenio 2010 CSD2008-0037 (CUP), Consolider-Ingenio 2010 CSD2007- 00042 (CPAN), and under contracts ref. FPA2008-03456, FPA2009-13697-C04-04; FCT(Lisbon) and FEDER under grant PTDC/FIS/103860/2008; the European Commission under the European Research Council T-REX Starting Grant ref. ERC-2009-StG-240054 of the IDEAS program of the 7th EU Frame-work Program; Director, Office of Science, Office of Basic Energy Sciences, of the US Department of Energy under contract no. DE-AC02-05CH11231. Part of these grants are funded by the European Regional Development Fund (ERDF/FEDER). J. Renner (LBNL) acknowledges the support of a US DOE NNSA Stewardship Science Graduate Fellowship under contract no. DE-FC52-08NA28752. F. I. acknowledges the support from the Eurotalents program. We are also grateful to our colleagues of the RD-51 collaboration for helpful discussions and encouragement. Finally, authors would like to acknowledge the use of Servicio General de Apoyo a la Investigación-SAI of the Universidad de Zaragoza and R. de Oliveira and his team at CERN for the manufacturing of the Micromegas readouts.